\documentclass[usenatbib]{emulateapj}
\usepackage{xspace}

\usepackage{hyperref}
\usepackage{cleveref}

\usepackage{epstopdf}
\usepackage{inputenc}
\usepackage[T1]{fontenc}
\usepackage{lmodern}
\usepackage{inputenc}
\usepackage{epsfig}
\usepackage{epstopdf}
\usepackage{amsmath}
\usepackage{lmodern}
\usepackage[T1]{fontenc}
\usepackage{natbib}
\usepackage{relsize}

\newcommand{\cotwo}{$^{12}$CO(2-1)~}
\newcommand{\coone}{$^{12}$CO(1-0)~}
\newcommand{\cothree}{$^{12}$CO(3-2)~}
\newcommand{\cotwop}{$^{12}$CO(2-1)}
\newcommand{\coonep}{$^{12}$CO(1-0)}
\newcommand{\cothreep}{$^{12}$CO(3-2)}
\newcommand{\kmsp}{km~s$^{-1}$}
\newcommand{\kms}{km~s$^{-1}$~}

\begin{document} 

\title{The Molecular Baryon Cycle of M~82} 
\author{John Chisholm\altaffilmark{1} \& Satoki Matsushita\altaffilmark{2}}
\shorttitle{The Molecular Baryon Cycle of M~82}
\shortauthors{CHISHOLM \& MATSUSHITA}
\altaffiltext{1}{Astronomy Department, University of Wisconsin,
Madison, 475 N. Charter St., WI 53711, USA}
\altaffiltext{2}{Academia Sinica Institute of Astrophysics and Astronomy, 11F of Astronomy-Mathematics Building, AS/NTU. No.1, Sec. 4, Roosevelt Rd, Taipei 10617, Taiwan, R.O.C}

\begin{abstract}
Baryons cycle into galaxies from the inter-galactic medium, are converted into stars, and a fraction of the baryons are ejected out of galaxies by stellar feedback. Here we present new high resolution (3\farcs9; 68~pc) \cotwo and \cothree images that probe these three stages of the baryon cycle in the nearby starburst M~82. We combine these new observations with previous \coone and [Fe~{\sc II}] images to study the physical conditions within the molecular gas. Using a Bayesian analysis and the radiative transfer code RADEX, we model molecular Hydrogen temperatures and densities, as well as CO column densities. Besides the disc, we concentrate on two regions within the galaxy: an expanding super-bubble and the base of a molecular streamer. Shock diagnostics, kinematics, and optical extinction suggest that the streamer is an inflowing filament, with a molecular gas mass inflow rate of 3.5~M$_\odot$~yr$^{-1}$. We measure the molecular gas mass outflow rate of the expanding super-bubble to be 17~M$_\odot$~yr$^{-1}$, 5 times higher than the inferred inflow rate, and 1.3 times the star formation rate of the galaxy. The high mass outflow rate and large star formation rate will deplete the galaxy of molecular gas within eight million years, unless there are additional sources of molecular gas.
\end{abstract}


\section{Introduction}
\label{intro}
Galaxies are not closed systems. Galaxies lose gas by converting a small portion (typically $\sim$1\%) of the gas into stars \citep{kennicutt, bigiel, leroy08}. A fraction of these newly formed stars are high-mass stars that emit high-energy photons, cosmic rays, and eventually explode as supernovae, which accelerates gas out of the star forming regions and into a galaxy-scale outflow \citep{chevalier, heckman90, heckman2000, veilleux05}. A portion of this outflow may escape the galactic potential \citep{heckman90, martin, rupkee, chisholm}, but lower velocity gas recycles back into the galaxy as a galactic fountain \citep{shapiro}. This migration of metal enriched gas shapes the mass-metallicity relation \citep{tremonti04, finlator, zahid}, and enriches the circum-galactic medium with metals \citep{tumlinson, werk, peeples}.

Meanwhile, galaxies gain gas through accretion from the circum-galactic medium \citep{katz03, keres05, keres09, dekel09}, as either cold filaments, or hot spherical accretion \citep{keres05, dekel09}. Accretion replenishes the gas lost through galactic outflows and star formation, promoting future star formation. Without accretion, galaxies consume their gas within a billion years \citep{leroy08, genzel}. 

The baryon cycle is the story of how galaxies acquire, store, and expel baryons. This cycle establishes the star formation history of the universe \citep{oppenheimer, hopkins}, the efficiency of star formation \citep{hopkins}, the gas fractions of galaxies \citep{dave11}, and the ratio of baryonic to non-baryonic matter within galaxies \citep{moster}.

Mergers and interactions speed up the baryon cycle. During the merger process, gravitational torques strip and compress the gas, increasing the inflow rate \citep{dimatteo, springel05, hopkins06}, the star formation efficiency \citep{mihos, saintonge}, and the mass outflow rate and velocity of the outflow \citep{hopkinswinds13, chisholm}. By increasing the amount of inflowing and outflowing material, the baryon cycle is easier to detect in merging and interacting galaxies.

\begin{deluxetable*}{ccccc}
\tablewidth{0pt}
\tablecaption{Calibration Properties}
\tablehead{
\colhead{Line} &
\colhead{Dates} & 
\colhead{Bandpass} &
\colhead{Flux} &
\colhead{Gain}
}
\startdata
\cotwo & 2004 Mar 7 & Callisto & Callisto & 0721+713 \\
       & 2004 Mar 12 & Ganymede and 3C279 & Ganymede & 0927+390 \\
\cothree & 2005 Feb 25 & Callisto & Callisto & 0721+713 and 0958+655
\enddata
\tablecomments{Table of the calibrators we use for the data reduction. \cotwo has two days of observations, and we list the sources used for both calibrations.}
\label{tab:obs}
\end{deluxetable*}

M~82 is a spectacular nearby (3.6~Mpc; \citealt{freedman}) starburst that is interacting with the massive spiral M~81 \citep{yun}. Due to its proximity, M~82 is close enough to map the full baryon cycle. Near the center of M~82, a starburst with a star formation rate of 13~$M_\odot$~yr$^{-1}$ \citep{forster03} drives a galactic outflow \citep{lynds, bland88, shopbell} that extends more than 2~kpc from the starburst \citep{bland88, shopbell, leroy15}. This galactic outflow has been observed in hot X-ray emitting plasma \citep{griffiths, strickland2004, strickland09}, ionized H$\alpha$ emission \citep{lynds, shopbell, ohyama, westmoquette2009a, sharp10}, and cold molecular gas \citep{weiss99, weiss2001, weiss2005,  matsushita2000, matsushita2005, mao2000, walter, keto05, salak13, leroy15}. Since stars form out of molecular gas, the cycle of cold gas into and out of the galaxy characterizes the near-future star formation within galaxies.

Here we trace the baryon cycle of diffuse molecular gas using three $^{12}$CO emission lines. In \autoref{sma} we describe new high resolution \cotwo and \cothree images taken with the Sub-Millimeter Array (SMA). We then include previous \coone observations (\autoref{nma}),  and infrared [Fe~{\sc II}] emission maps (\autoref{fe2}) to characterize the physical conditions within the galaxy. The $^{12}$CO intensity, velocity, and channel maps are first presented in \autoref{maps}, and then we use a Bayesian analysis, with the radiative transfer code RADEX, to model the temperatures and densities of the molecular gas (\autoref{prop}).

We focus on three important physical regions within M~82: the disc (\autoref{disc}), the expanding super-bubble (\autoref{bubble}), and the base of a molecular streamer (\autoref{s2}). Using the derived densities, we measure the masses of each component, and find that the super-bubble has a molecular mass outflow rate of 17~M$_\odot$ yr$^{-1}$. In \autoref{s2}, the kinematics, shock diagnostics, and optical extinction suggest that the molecular streamer is an inflowing filament, with a molecular inflow rate of 3.5~M$_\odot$ yr$^{-1}$. Finally, we explore the molecular baryon cycle within M~82, finding that the star formation and outflow will consume the molecular gas in eight million years, unless there are an extra sources of molecular gas (\autoref{baryon}).

Throughout this paper we assume that M~82 is at a distance of 3.6~Mpc \citep{freedman}, and that 1\rq{}\rq{} corresponds to 17.5~pc.

\section{DATA AND OBSERVATIONS}
\label{observations}
Here we describe the individual observations of M~82. We first discuss the new \cotwo and \cothree observations (\autoref{sma}), and then include the \coone observations from \citet{matsushita2000} (\autoref{nma}). In \autoref{fe2obs} we introduce  the [Fe~{\sc II}] images, which are important shock diagnostics.

\subsection{SMA Observations}
\label{sma}

\cotwo observations were taken with the Sub-Millimeter Array \citep[SMA;][]{ho2004} on March~7th and 12th, 2004; while \cothree observations were taken on February~25th, 2005 (see \autoref{tab:obs}). The primary beam sizes of the \cotwo and \cothree observations are $55''$ and $36''$, respectively. We observe \cotwo with one pointing, and mosaic three \cothree  pointings to give a similar field of view. The \cotwo beam is centered on the central CO peak of M~82 \citep{shen1995}. The middle \cothree beam is also centered on the CO peak, with the other two pointings separated by $18''$ (i.e., Nyquist sampled) east and west along the major axis of the disc at a position angle of 75$^\circ$.

For the data calibration, we use the software package MIR, adopted for SMA\footnotemark[3]. \footnotetext[3]{\url{http://www.cfa.harvard.edu/~cqi/mircook.html}}
We flag and correct spikes in the system temperature, and make passband calibrations for the phase, amplitude, gain, and then flux calibrate the images using the sources listed in \autoref{tab:obs}. 

We then import the data into the Common Astronomy Software Applications package \citep[CASA;][]{mcmullin2007} and combine the individual observations. Using the dirty images, we find line emission in channels between $-195$~km~s$^{-1}$ and $+150$~km~s$^{-1}$ (the velocity zeropoint here is the local standard of rest velocity for M~82, 225~km~s$^{-1}$), and then subtract the continuum in the UV plane. We clean the continuum subtracted images to the RMS of the dirty image, producing images that over-sample the beam (pixel sizes of 0\farcs3 for \cotwop, and 0\farcs2 for \cothreep) and have a velocity resolution of 5~km~s$^{-1}$ to match the \coone observations below. We make a primary beam correction, and integrate intensities that are significant at greater than the $5\sigma$ level to produce zeroth moment maps. The original spatial resolutions are $3\farcs88 \times 3\farcs32$ at a position angle of $-26^\circ$ for \cotwop, and $2\farcs76 \times 2\farcs23$ at $-179^\circ$ for \cothreep.

 Since interferometers incompletely sample the UV plane, we correct for the missing flux with an ad-hoc short-spacing correction. The short-spacing correction is made by convolving the zeroth moment maps to the spatial resolution of previous single dish observations \citep{mao2000, weiss2005}, and then comparing the intensities within six regions of our interferometric data to the single dish intensities. The average intensity difference is 49\% and 36\% for the \cotwo and \cothree observations, respectively. We multiply this intensity difference by the number of pixels per beam, and divide by the velocity range integrated over to create the zeroth moment map. The short-spacing correction is then added to the individual channels. This short-spacing correction only corrects for the average short-spacing. A more robust treatment of the short-spacing corrections would use single-dish data to account for spatial variations in the UV plane coverage. In \autoref{maps} we compare individual points in our line ratio maps to single-dish maps, and assign extra uncertainty to the line ratios to account for point-to-point variations. We recreate the zeroth moment maps, convolve the \cothree map to the resolution of the \cotwo map, and convert the intensity into a brightness temperature.

\begin{figure*}
\includegraphics[width = \textwidth]{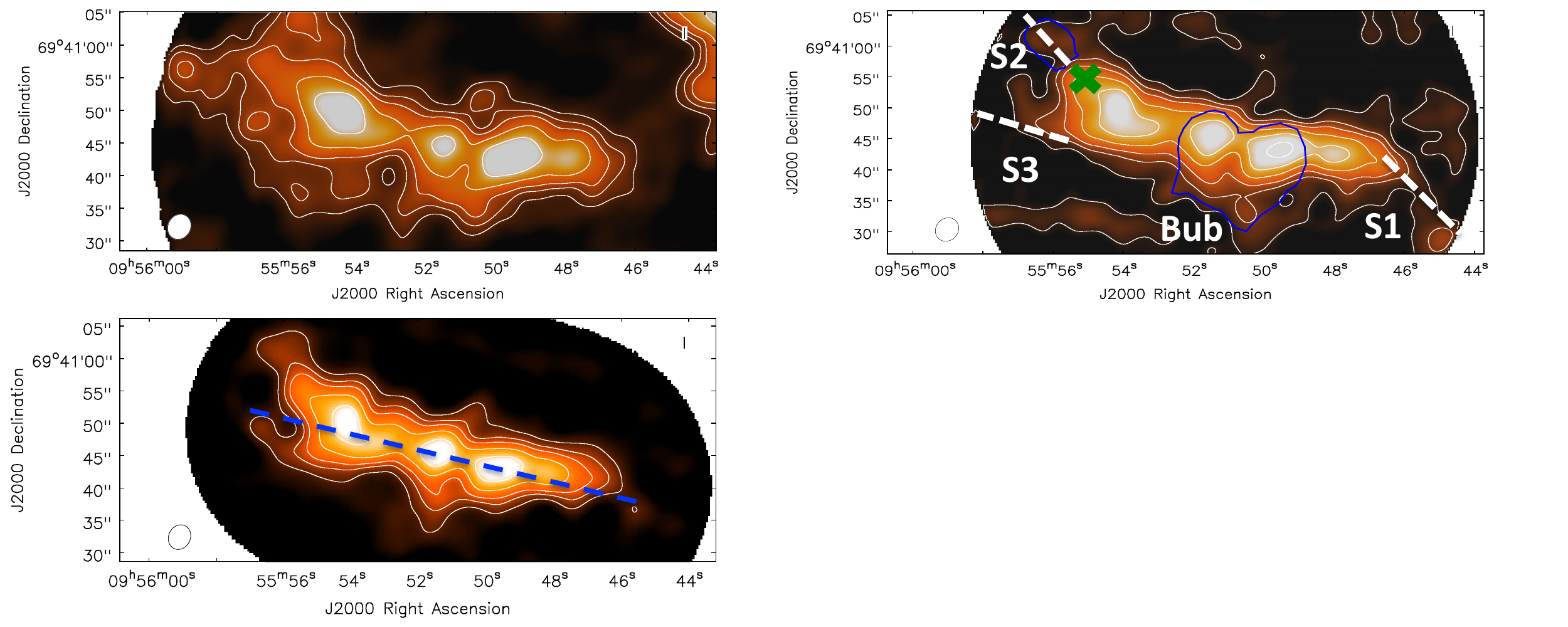}
\caption{The integrated intensity maps for \coone \citep[upper left panel;][]{matsushita2000, matsushita2005}, \cotwo (upper right panel) and \cothree (lower left panel). Numerous features from \citet{walter} and \citet{matsushita2000} are marked on the \cotwo map. These features include the western spur (S1), the northeastern spur (S2), the southeastern spur (S3), and the expanding super-bubble (bub; outlined in blue). The green X marks the approximate location of a strong knot of [Fe~{\sc II}] emission from \autoref{fig:fe2}. The blue regions illustrate the areas used to define the bubble (central structure; from the -85~\kms channel map), and the S2 streamer (northeastern rectangular shape; from the velocity map).  S1 and S3 are detected off the south west and east of the disc, respectively, in \cotwo and \cothreep, but the \coone is not deep enough to detect these features.  The blue dashed line on the \cothree map represents the position slice for the position-velocity diagrams. The beam size of $3\farcs88 \times 3\farcs32$ (68~pc$ \times $58~pc) at a position angle of -32$^\circ$ is indicated in the lower left corner of each panel. The contour levels are 450, 600, 750, 1050, 1500, and 2250~K~km~s$^{-1}$, and the estimated $1\sigma$ intensity errors are 103, 55, and 64 K~km/s for the \coonep, \cotwop, and \cothree maps, respectively.}
\label{fig:mom0}
\end{figure*}

\subsection{NMA Observations}
\label{nma}

The Nobeyama Millimeter Array (NMA) observed M~82 in \coone between November 1997 and March 1999, and the full reduction details are given in \citet{matsushita2000}. The NMA data has a velocity resolution of 5.2~km~s$^{-1}$, a beam size of $2\farcs8 \times 2\farcs3$ at $-50^\circ$, and a pixel size of $0\farcs25$. Similar to the SMA data above, a primary beam correction is made, and a zeroth moment map is created by integrating between $-180$ and $+150$ km~s$^{-1}$.  We make short-spacing corrections, similar to the \cotwo and \cothree maps, with a 9\% difference between the single dish data \citep{weiss2005} and the NMA data. Finally, the zeroth moment map is convolved to the resolution of the \cotwo map.

\subsection{HST [Fe~{\sc II}] Imaging}
\label{fe2obs}

We use archival Hubble Space Telescope [Fe~{\sc II}]~1.6~$\mu$m images to determine whether the gas is shocked \citep{alonso}. The NICMOS images have a field of view of 53\farcs8 by 66\farcs2, and a resolution of 0\farcs35, more than adequate to compare to the CO observations. The images are reduced using the NicRed package \citep{mcleod}, flux calibrated, and continuum subtracted using off-band F164N images \citep{alonso}.

\section{RESULTS}

Here we present the results of the CO and [Fe~{\sc II}] emission maps. We first introduce the CO intensity, velocity, position-velocity, channel, and line ratio maps (\autoref{maps}). We then model H$_2$ temperature and density maps using a Bayesian approach (\autoref{prop}). Finally we use the [Fe~{\sc II}] emission to locate shocked gas within M~82 (\autoref{fe2}).

\subsection{Moment, Channel, and Line Ratio Maps}
\label{maps}

\begin{figure*}
\includegraphics[width = \textwidth]{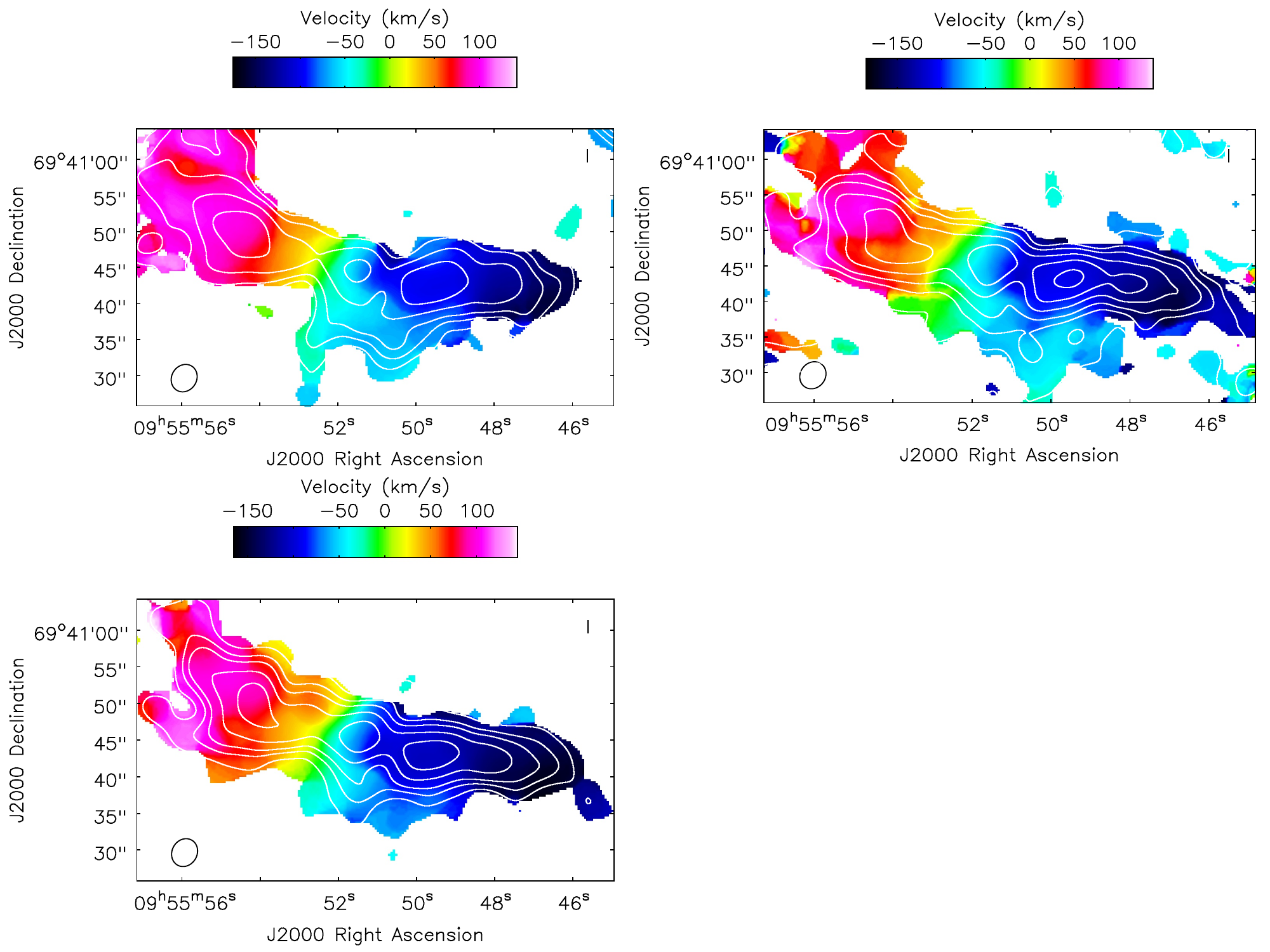}
\caption{Maps of the intensity weighted velocity relative to the local standard of rest (225~\kmsp) for \coone (top left panel), \cotwo (top right panel), and \cothree (bottom panel). Only pixels with 5$\sigma$ or greater detections are used to create the map. The S2 and bubble features are both kinematically distinct from the nearby disc (see the red and orange amongst pink for S2, and light blue amongst dark blue for the super-bubble). S2 is blueshifted by $37$~\kmsp, and the bubble is blueshifted by $50$~\kms relative to the rotating disc. The contours are from the zeroth moment maps (\autoref{fig:mom0}), and the beam size (68~pc$ \times $58~pc at the distance of M~82) is given in the lower left corner.}
\label{fig:mom1}
\end{figure*}

\begin{figure*}
\includegraphics[width = \textwidth]{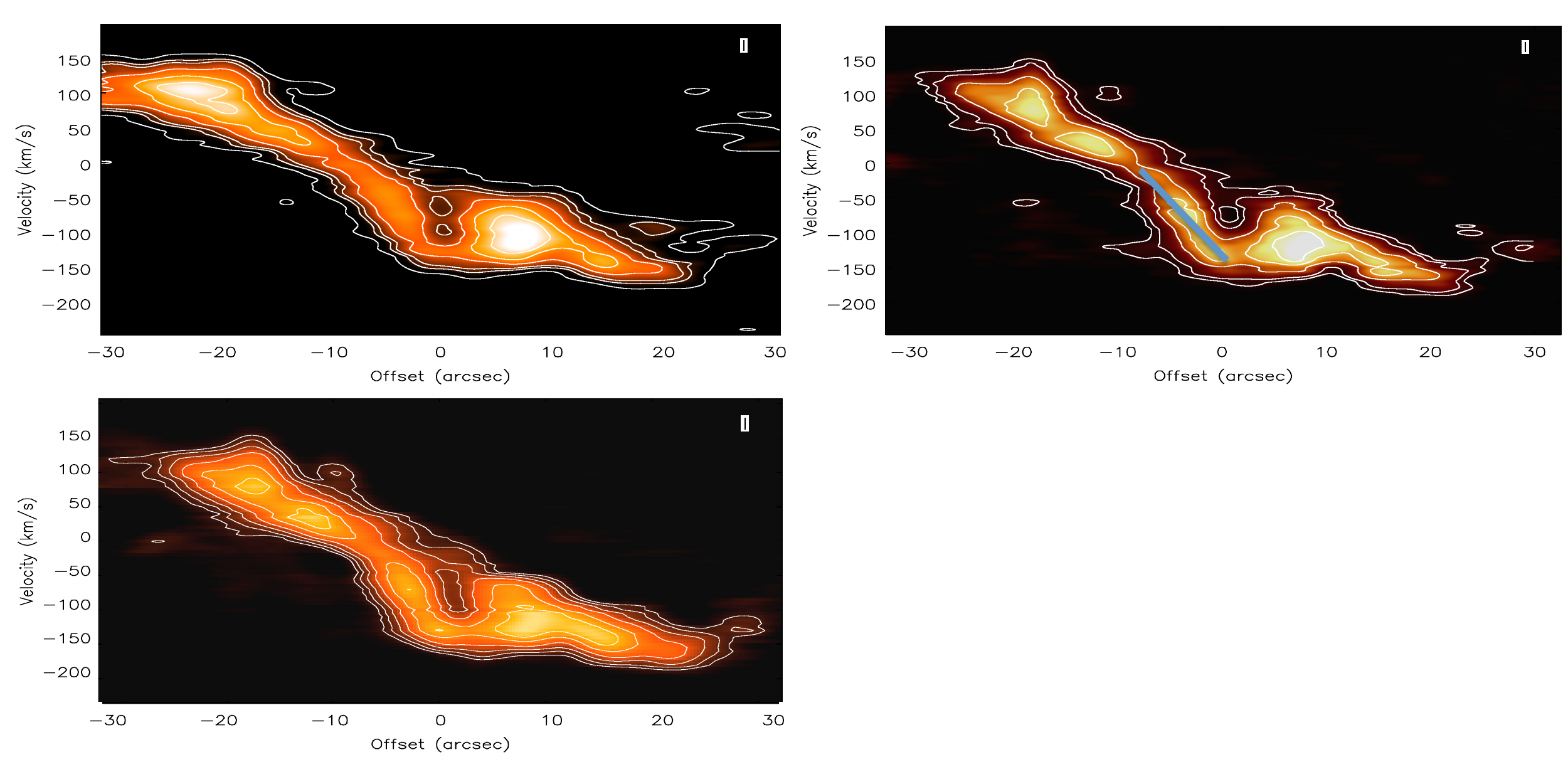}
\caption{The position-velocity diagram for \coone (top left panel), \cotwo (top right panel), and \cothree (bottom left panel) along the dashed blue line in \autoref{fig:mom0}. Zero velocity is defined as the velocity of the local standard of rest, or 225~\kmsp.  A deviation from rigid body rotation is seen near a position of $0''$ (9$^h$55$^m$51.0$^s$, +69$^\circ40'44.85''$), corresponding to the expanding super-bubble. The emission from the bubble is concentrated between velocities of $-10$~km~s$^{-1}$ and $-139$~km~s$^{-1}$. The velocity gradient used to calculate the mass outflow rate of the bubble is given as the solid light-blue line in the \cotwo panel (upper right; see \autoref{bubble}). The velocity resolution is 5~\kmsp. }
\label{fig:pv21}
\end{figure*}

A zeroth moment map measures the integrated intensity in each pixel, and the maps for the three CO emission lines are shown in \autoref{fig:mom0}. To measure the  standard deviation of the zeroth moment maps, we first measure the standard deviation in the individual channels, then multiply by the velocity resolution (5~\kmsp) and the square root of the number of channels. The zeroth moment $1\sigma$ errors are 103, 55, and 64~K~km~s$^{-1}$ for the  \coonep, \cotwop, and \cothree maps, respectively. Three distinct intensity peaks are seen \citep{shen1995, weiss2001, walter}, and four previously observed large-scale features are detected: three molecular spurs off the disc \citep[named S1, S2, S3;][]{walter}, and an expanding super-bubble \citep{weiss99, matsushita2000}.

The intensity weighted velocity maps (also called the first moment maps) are shown in \autoref{fig:mom1}. Galactic rotation is seen across the disc, with the velocity steadily increasing from east to west. The rotational center of the galaxy resides between the left two peaks in \autoref{fig:mom0}, while an evolved 10$^6$~M$_\odot$ super star cluster is between the right two CO intensity peaks \citep{matsushita2000}. 

Slicing the image cube along the dashed blue line in the bottom panel of \autoref{fig:mom0}, we produce the position-velocity (PV) diagrams for each transition (\autoref{fig:pv21}). A PV diagram plots the intensity at a particular position and velocity, and diagnoses the kinematics of the gas. The PV diagrams show a steady decrease in velocity with increasing position (i.e., rigid rotation). Near the $0''$ position the CO deviates from the rigid rotation. \citet{matsushita2000} attribute this abrupt change in velocity  to an expanding molecular super-bubble.

\begin{figure}
\includegraphics[width = .5\textwidth]{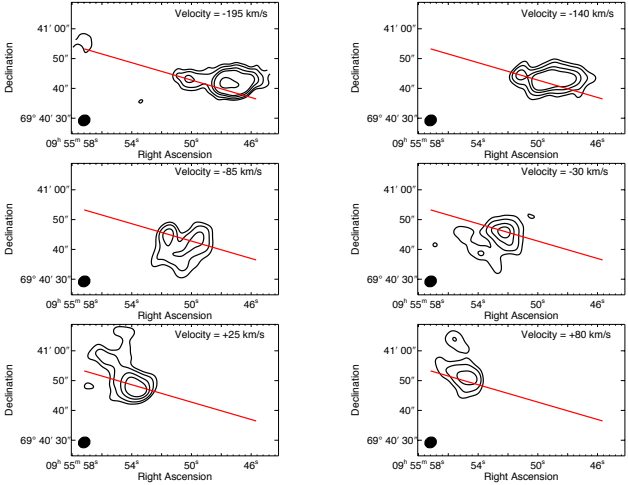}
\caption{Channel maps for the \coone emission. The six velocity channels are separated by 55~\kmsp, with median velocity given in the top right corner of each map. Zero velocity is defined as the velocity of the local standard of rest (225~\kmsp). The contour levels represent a 7, 10, 15, 20, 40$\sigma$ detection significance. The red-line is the line used to produce the PV diagrams (see \autoref{fig:mom0}), and represents the major axis of the galaxy}.  The beam size is given in the lower left corner, and is \textbf{68~pc$ \times 58$~pc at the distance of M~82. }
\label{fig:channels1}
\end{figure}

\begin{figure}
\includegraphics[width = .5\textwidth]{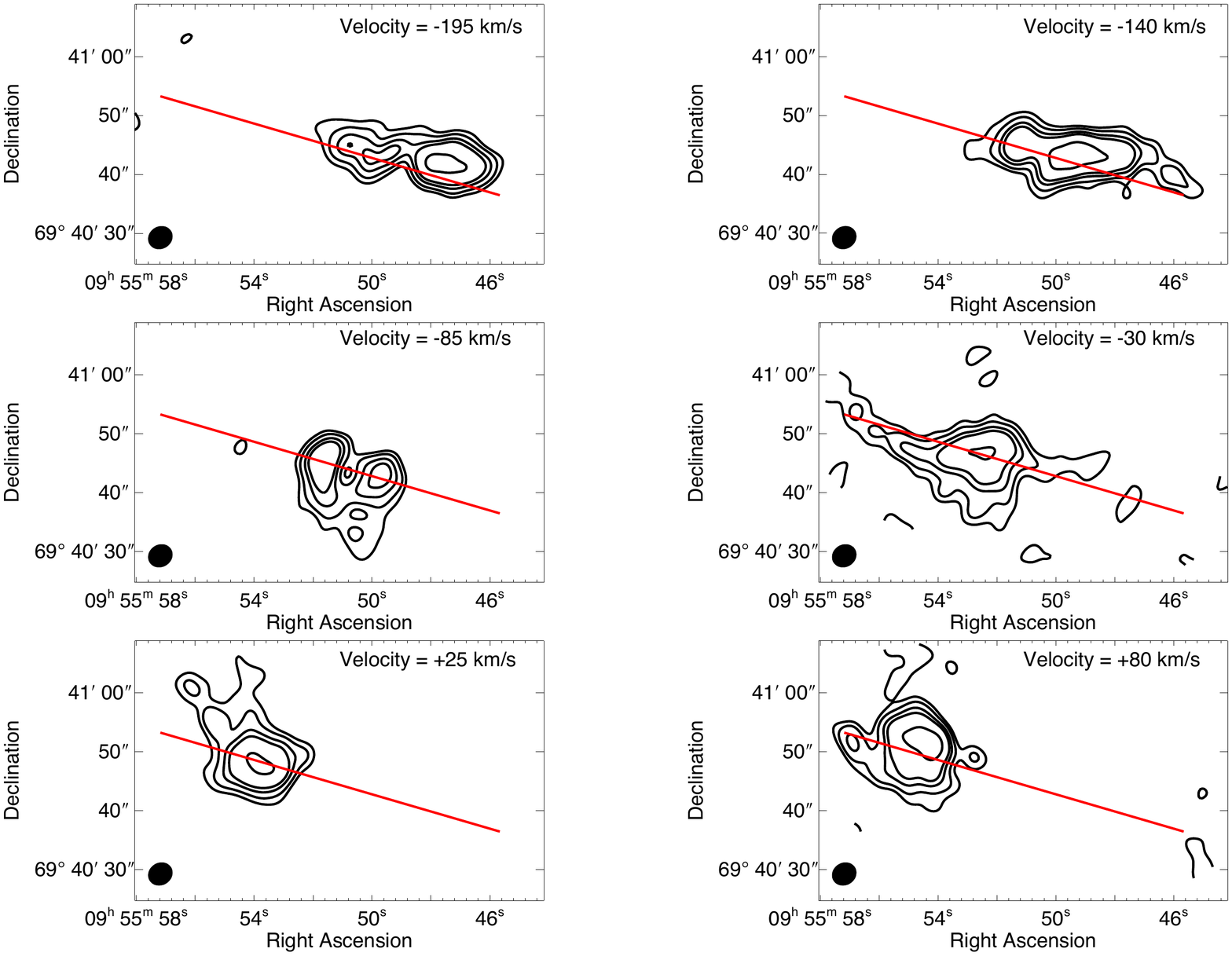}
\caption{Same as \autoref{fig:channels1} but for the \cotwo emission.}
\label{fig:channels2}
\end{figure}

\begin{figure}
\includegraphics[width = .5\textwidth]{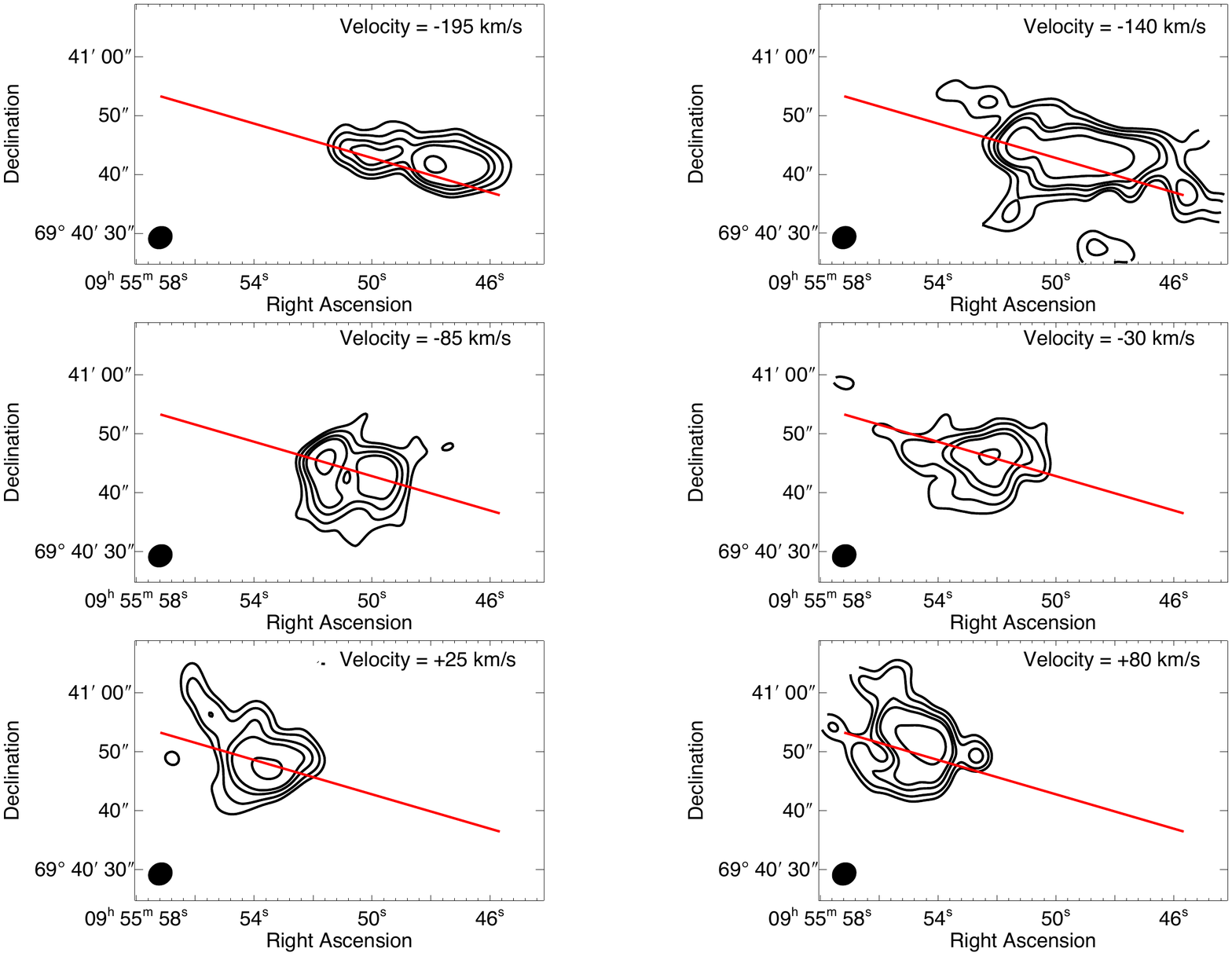}
\caption{Same as \autoref{fig:channels1}, but for  the \cothree emission.}
\label{fig:channels3}
\end{figure}

We also use the channel maps -- or the intensity within a given velocity interval -- to study the distribution of CO gas in velocity space. \Cref{fig:channels1,fig:channels2,fig:channels3} show the channel maps for \coonep, \cotwop, and \cothreep, respectively. We define six channels, separated by 55~\kmsp, with median velocities given in the upper right corners of each map. The $-195$~\kmsp, $-140$~\kmsp, $-30$~\kmsp, and $+80$~\kms channels largely contain disc emission. The $-85$~\kms channel (middle left panel) has emission from the super-bubble, and the area defined by this feature is shown by the blue circle in the upper right panel of \autoref{fig:mom0}. Moreover, we detect the S2 streamer from \citet{walter} in the upper left corner of the $+25$~\kmsp channel (lower left panel). The \cotwo and \cothree lines show more streamers: S1 is seen in the bottom right of the  $-140$~\kms channel and in the lower right of the \cothree PV diagram. Additionally, S3 is seen in the bottom left of the $+80$~\kms channel (bottom left panel). Unfortunately, the sensitivity of the \coone observations from \citet{matsushita2000} do not afford detections of these diffuse features.

\begin{figure}
\includegraphics[width = .5\textwidth]{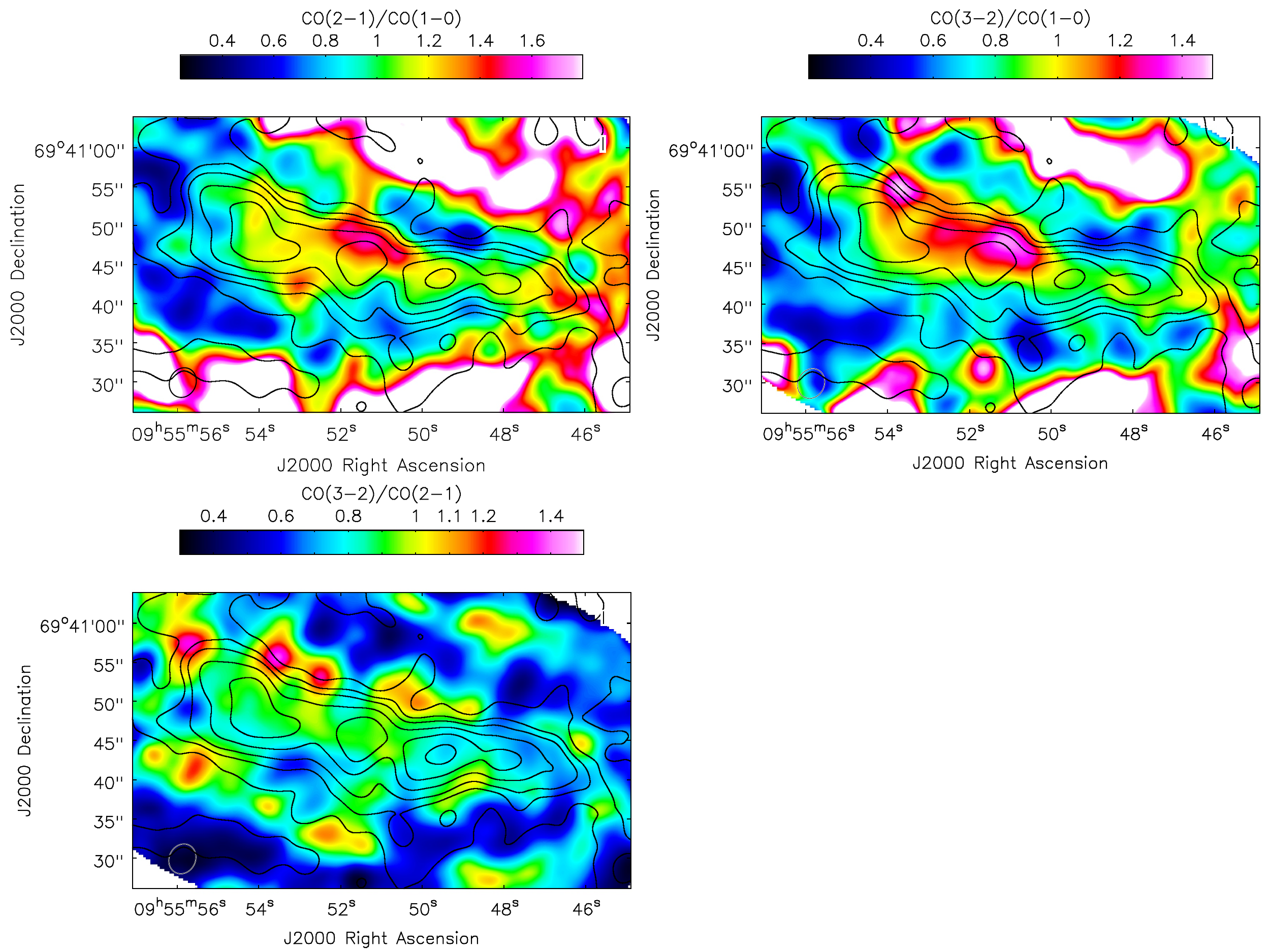}
\caption{The CO line ratio maps used to model the molecular temperatures and densities. Only 1$\sigma$ detections are included in the map. The upper left panel is the \cotwop/\coone ratio, the upper right panel is the \cothreep/\coone ratio, and the lower left panel is the \cothreep/\cotwo ratio. \autoref{tab:linerat} tabulates the ratios for five different regions, with points in the disc, the super-bubble, and S2. The black contours are the \cotwo contour levels from \autoref{fig:mom0}, and the beam size \textbf{(68~pc$ \times $58~pc)} is shown in the lower left corner.}
\label{fig:linerat}
\end{figure}

In \autoref{prop} we use the radiative transfer code RADEX and the line ratios to model the temperatures and densities of the molecular gas in M~82. \autoref{fig:linerat} shows the \cotwop/\coonep, \cothreep/\coonep, and the \cothreep/\cotwo line ratio maps. Additionally, in \autoref{tab:linerat} we tabulate the CO emission lines for five regions discussed in \autoref{discussion}. These line ratios provide the basis to model the physical properties of the molecular gas. To calculate the errors on these line ratio maps we must consider the dominant uncertainties in our calibration procedure, which are the calibration errors (short-spacing corrections), not the statistical uncertainties. Therefore, for the line ratio errors we use a calibration uncertainty of 15\% and an uncertainty due to our ad hoc short-spacing correction. We make a point-by-point comparison between single dish data from \citet{weiss2005}, \citet{wilson12}, and \citet{leroy15} to our line ratios and find a 20\% variation. We conservatively attribute this 20\% variation to the ad-hoc prescription for the short-spacing correction. We then add the short-spacing error in quadrature with the 15\% calibration uncertainty to find an uncertainty of 25\% on the individual line ratios. This is the uncertainty that we model the temperatures and densities with, below.

\begin{deluxetable}{ccc}
\tablewidth{0pt}
\tablecaption{RADEX Model Parameter Ranges}
\tablehead{
\colhead{Parameter} &
\colhead{Range} & 
\colhead{Step Size}  \\
}
\startdata
Temperature [K]                &  15 -- 295 & 10 \\
log($n_\text{H2}$) [cm$^{-3}$] &  1 --   5 &  0.1 \\
log($N_\text{CO}$) [cm$^{-2}$] & 17 --  23 &  0.1 \\
\enddata
\tablecomments{Grid of parameters used to create the RADEX models.}
\label{tab:mod}
\end{deluxetable}

\begin{deluxetable*}{ccccccccc}
\tablewidth{0pt}
\tablecaption{Line Ratios}
\tablehead{
\colhead{Feature} &
\colhead{R.A. (2000)} & 
\colhead{Decl. (2000)} &
\colhead{\cotwop/\coone} &
\colhead{\cothreep/\coone} &
\colhead{\cothreep/\cotwo} & 
\colhead{$\tau_{10}$} & 
\colhead{$\tau_{21}$} & 
\colhead{$\tau_{32}$}\\
\colhead{} &
\colhead{(hh:mm:ss)} & 
\colhead{(hh:mm:ss)} &
\colhead{} &
\colhead{} &
\colhead{} &
\colhead{}
}
\startdata
Disc & 09:55:52.4 & +69:40:46.08 & $1.17 \pm 0.29$ & $1.10 \pm 0.28$ & $0.94 \pm 0.24$ & 10 & 36 & 73\\ 
Shocked Bubble & 09:55:51.0 & +69:40:44.82 & $0.62 \pm 0.16$ & $0.80 \pm 0.2$ & $1.30 \pm 0.33$ & 10 & 39 & 79 \\ 
Outer Bubble & 09:55:50.9 & +69:40:36.62 & $0.38 \pm 0.10$ & $0.33 \pm 0.08$ & $0.89 \pm 0.22$ & 33 & 120 & 226 \\ 
Shocked S2 & 09:55:55.4 & +69:40:55.90 & $0.80 \pm 0.2$ & $0.78 \pm 0.20$ & $0.98 \pm 0.25$ & 58 & 194 & 327 \\ 
S2 Body & 09:55:56.2 & +69:41:00.00 & $0.59 \pm 0.15$ & $0.57 \pm 0.14$ & $0.96 \pm 0.24$ & 32 & 105 & 176
\enddata
\tablecomments{Table of CO line ratios for five points in M~82. The errors on the line ratios are 25\%, as outlined in \autoref{maps}. The individual points correspond to: the 2~$\mu$m peak position \citep[disc;][]{lester}, a shocked portion of the super bubble, a position further out in the southern bubble (outer bubble), the location of [Fe~{\sc II}] emission at the intersection of S2 and the disc (shocked S2), and a position in the body of S2.  The ratios for the disc are calculated using the integrated intensity maps, while the bubble and S2 ratios are calculated only with the $-85$~\kms and $+25$~\kms channel maps. The final three columns give the RADEX modeled optical depths for the \coonep, \cotwop, and \cothree transitions, respectively.}
\label{tab:linerat}
\end{deluxetable*}

\subsection{Density and Temperature Modeling}
\label{prop}

To model the molecular gas temperatures and densities, we compare the ratios of the \coonep, \cotwop, and \cothree intensities to theoretical ratios from RADEX \citep{radex}. RADEX is a one-dimensional, non-LTE, radiative transfer code that calculates the molecular line ratios using an escape probability formulation. RADEX requires inputs of temperature, H$_2$ density ($n_\text{H2}$), CO column density ($N_\text{CO}$), and the FWHM of the CO line.  We fit the line widths at each pixel for each transition, and find a median FWHM of 89~\kmsp. By using the FHWM of the entire galaxy for the RADEX modeling, we assume that the CO emitting gas is in a single zone, or the entire galaxy is treated as a single cloud of molecular gas. We make this assumption to simplify the Bayesian parameter grid below.

\begin{figure}
\includegraphics[width = 0.5\textwidth]{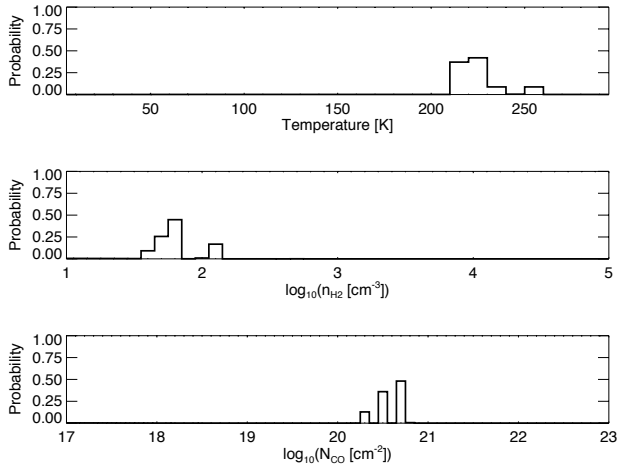}
\caption{Example probability density functions (PDFs) for a single pixel within the -85~\kms channel (the hot region of the super-bubble). Each panel gives the normalized probability of the model given the observed line ratios. The different panels correspond to the PDFs for the temperature (top panel), the molecular Hydrogen number density (middle panel), and CO column density (bottom panel). The temperature, Hydrogen density (log($n_\mathrm{H2}$)), and CO column density (log($N_\mathrm{CO}$)) derived from these PDFs are $222\pm 22$~K, $1.79\pm0.18$~dex, and $20.50\pm0.50$~dex, respectively.}
\label{fig:pdfs}
\end{figure}

We then create a large grid of RADEX models using the parameters listed in \autoref{tab:mod}, and tabulate the predicted line ratios for each input parameter. The RADEX grid is centered on the warm diffuse molecular temperatures and densities found in \citet{weiss2005}, but the grid also allows for very cold dense clouds and the 272~K molecular phase  from H$_2$ rotational transitions \citep{beirao}.  We use a Bayesian analysis to calculate the probability of each RADEX model given the observed line ratios (\cotwop/\coonep, \cothreep/\cotwop, and \cothreep/\coonep), where only locations detected at greater than 1$\sigma$, for all three lines, are used. We create probability density functions (PDFs) for each marginalized parameter by assuming a flat prior, such that each value is equally likely, and compute the likelihood function as \citep{brinchmann, kauffmann, stats}:
\begin{equation}
L \propto \mathrm{exp}(-\chi^2/2).
\end{equation}
The intensity errors (used in the $\chi^2$ function) are 25\% from the calibration and short-spacing correction (see \autoref{maps}). The likelihood functions are then normalized, and marginalized over nuisance parameters to produce PDFs for $n_\text{H2}$, temperature, and $N_\text{CO}$ individually. The PDFs are typically narrowly peaked, and clustered near a single value (see \autoref{fig:pdfs}), although the temperature PDFs occasionally are broader. The expectation values and standard deviations are calculated from the PDFs. \autoref{fig:pdfs} gives a representative example of the three PDFs, the expectation values, and standard deviations from a single pixel within the hot super-bubble region of the $-85$~\kms channel. 

\begin{figure*}
\includegraphics[width = \textwidth]{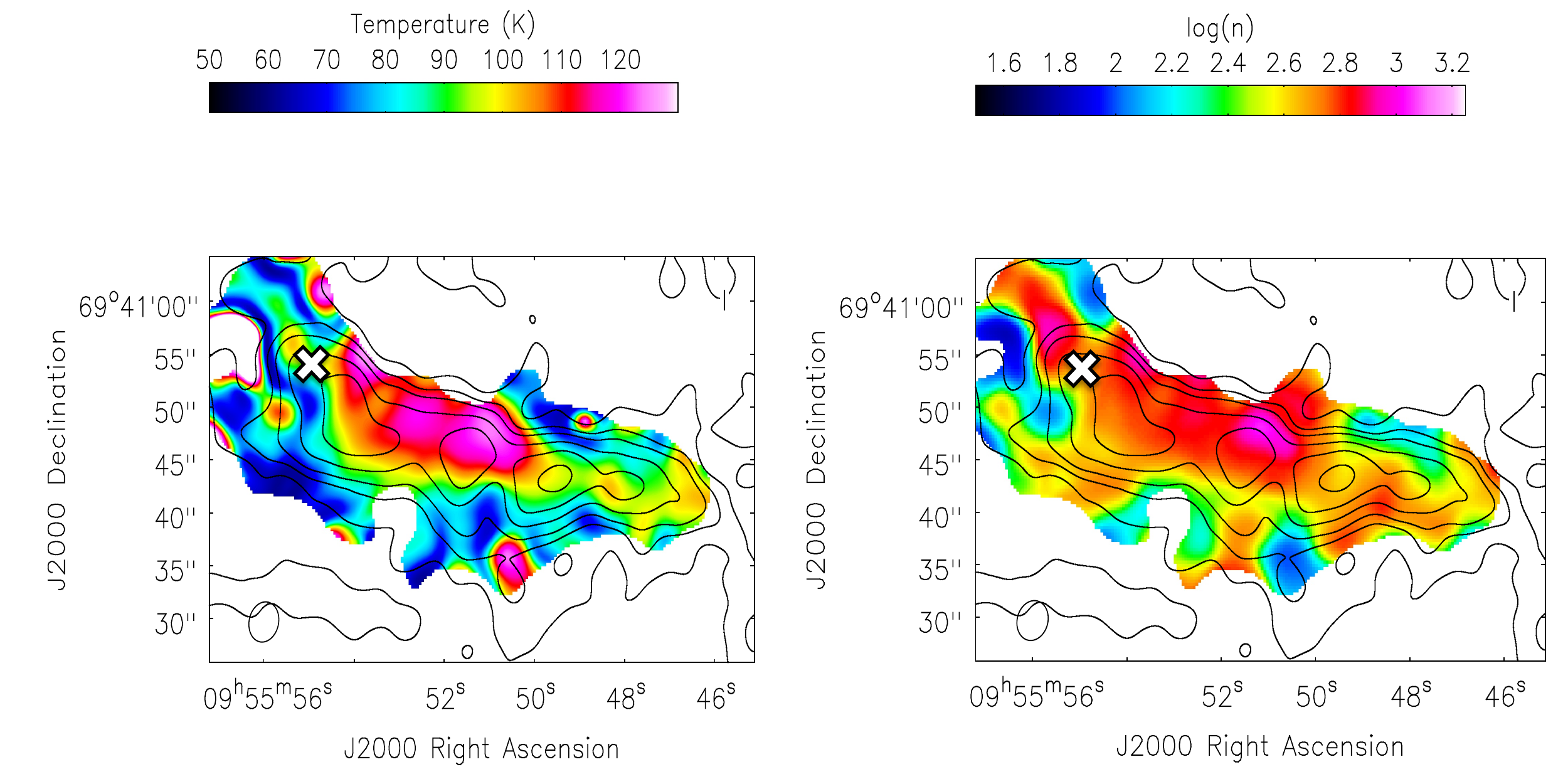}
\caption{Maps of the expectation values for the H$_2$ temperature (left; given in K) and density (right; given in log(cm$^{-3}$)). The \cotwo contour map from \autoref{fig:mom0} is overlaid in black to give a sense of position. The white X's mark the approximate location of the [Fe~{\sc II}] emission from \autoref{fe2}.}
\label{fig:properties}
\end{figure*}

\begin{figure*}
\includegraphics[width = \textwidth]{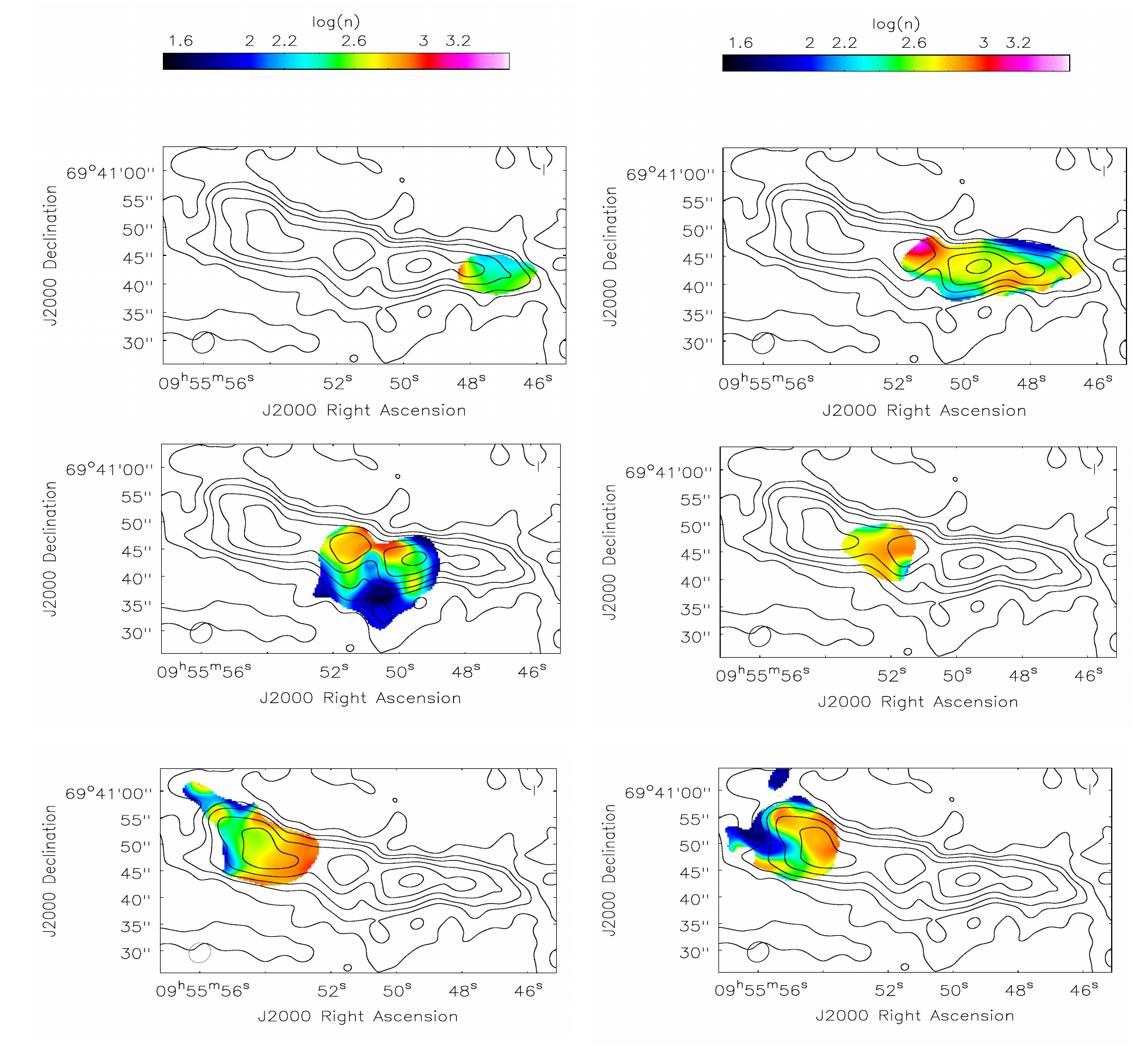}
\caption{Density channel maps with the \cotwo contours from \autoref{fig:mom0} overlaid to give a sense of position. The density is only displayed if the three CO lines are detected at more than a 1$\sigma$ significance level. The six maps are the same velocities as \autoref{fig:channels1}: $-195$~\kmsp, $-140$~\kmsp, $-85$~\kmsp, $-30$~\kmsp, $+25$~\kmsp, and $+80$~\kmsp. S2 is seen in the $+25$~\kms channel (lower left), and the bubble region is seen in the $-85$~\kms channel (middle left). }
\label{fig:chanden}
\end{figure*}

\begin{figure*}
\includegraphics[width = \textwidth]{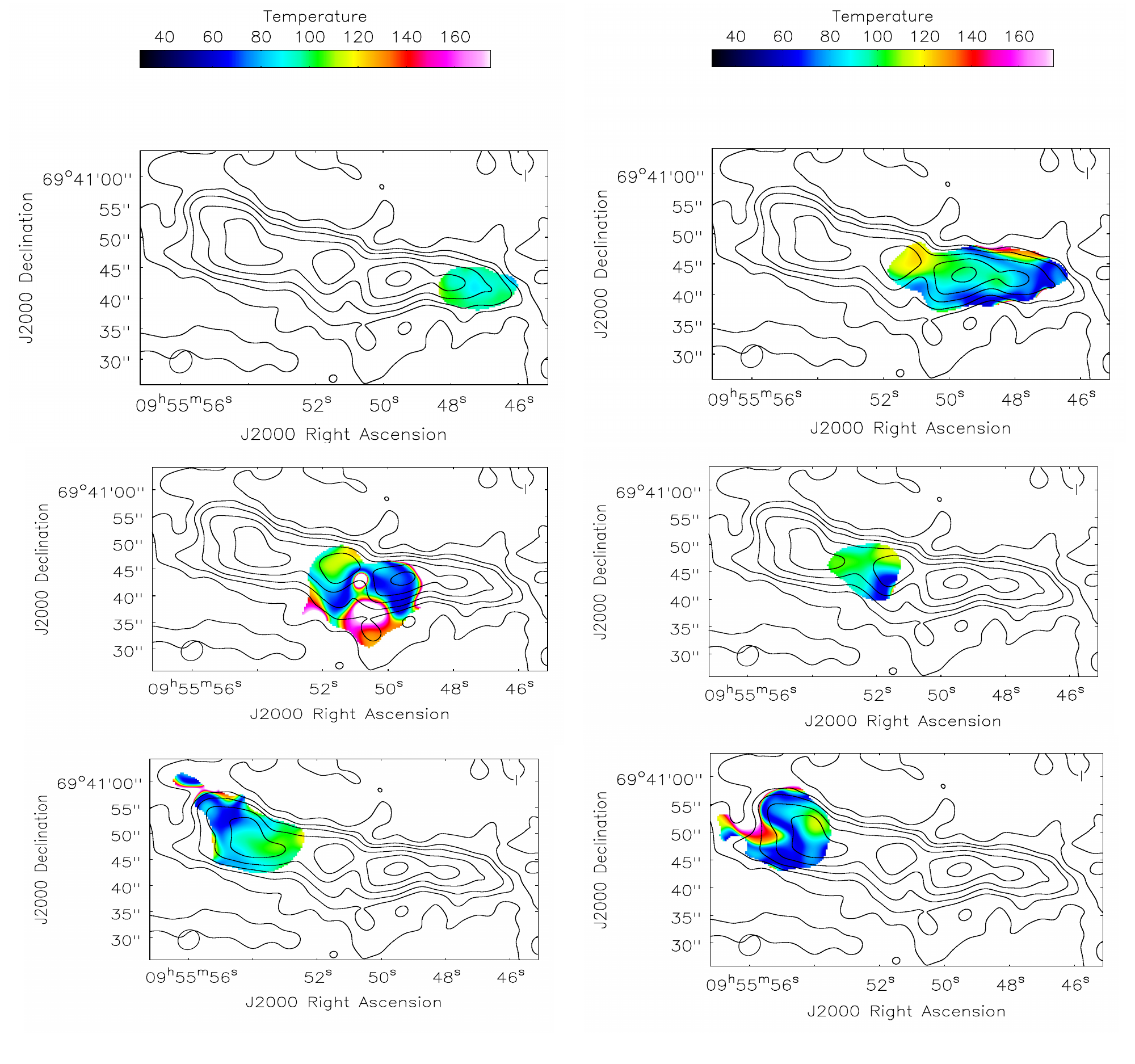}
\caption{Same as \autoref{fig:chanden}, but for the H$_2$ temperature.}
\label{fig:chantemp}
\end{figure*}

We plot expectation value maps for the temperature and $n_\text{H2}$ over the entire field of view in \autoref{fig:properties}. The log($n_\text{H2}$) ranges between 1.8 and 3.1~dex, and temperatures range between 61 and 175~K. The median individual standard deviations of the temperatures and densities are 70~K and 0.66~dex, respectively. \Cref{fig:chanden,fig:chantemp} show the temperatures and densities modeled from the channel maps, showing the conditions within specific features, like the super-bubble and S2.

\begin{figure}
\includegraphics[width = .5\textwidth]{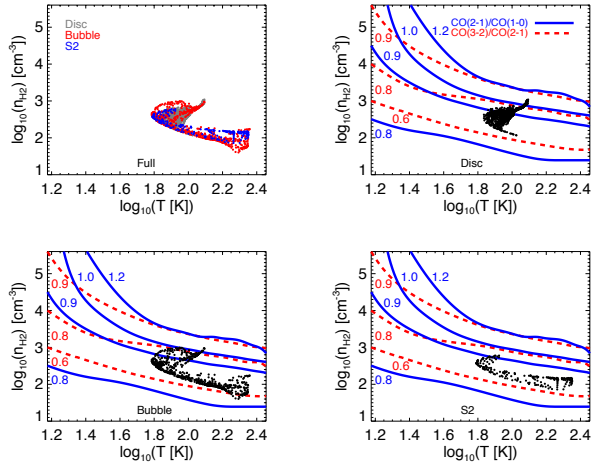}
\caption{A random sampling of pixels in the temperature-density plane for the full field of view (upper left), disc (upper right), bubble (lower left) and the S2 region (lower right). In the full field of view, we have color-coded the pixels by whether they are in the disc (gray), the super-bubble (red), or the S2 molecular streamer (blue). Lines of constant \cotwop/\coone (blue lines) and \cothreep/\cotwo (red dashed lines) ratios are shown in the single component panels, created using a constant log($N_\mathrm{CO}$[cm$^{-2}$]) = 19. The constant ratio lines are labeled near the lowest temperature of each curve.}
\label{fig:phasespace}
\end{figure}

\begin{figure}
\includegraphics[width = .5\textwidth]{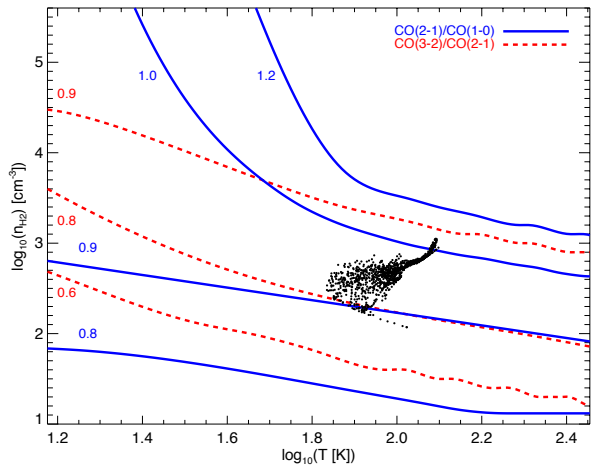}
\caption{Same as the disc panel from  \autoref{fig:phasespace} (upper right), but for a log($N_\mathrm{CO}$) = 19.3 (twice the column density of \autoref{fig:phasespace}).  Generally, increasing log($N_\mathrm{CO}$) shifts these curves to lower densities, but the \cotwop/\coone ratio evolves more rapidly than the \cothreep/\cotwop. $N_\mathrm{CO}$ can then be estimated by marginalizing over the density and temperature.}
\label{fig:phasespacehigh}
\end{figure}

Additionally, we show the relationship between the modeled temperatures and densities (\autoref{fig:phasespace}). We over-plot curves of constant \cotwop/\coone and \cothreep/\coone line ratios from the RADEX models, using a constant log($N_\mathrm{CO}$) of 19~dex. In \autoref{fig:phasespacehigh} we show the curves corresponding to the same constant line ratios, but for a log($N_\mathrm{CO}$) of 19.3, twice the column density in \autoref{fig:phasespace}. While both the \cotwop/\coone and \cothreep/\cotwo change by changing $N_\mathrm{CO}$, the \cotwop/\coone moves to lower temperatures and densities more rapidly than the \cothreep/\cotwop. This relation allows for low \cotwop/\coone yet high \cothreep/\cotwo ratios seen in the outer bubble and the S2 body to correspond to low N$_\mathrm{CO}$ values (see \autoref{tab:linerat}). This allows for the $N_\mathrm{CO}$ values to be modeled by marginalizing over the PDFs.

The RADEX modeling also provides estimates of the optical depth for each transition. In \autoref{tab:linerat} we show the optical depth for each transition in five regions that are discussed below. These optical depths are moderately high and consistent with optical depths derived from single-dish data \citep{leroy15}. Observations of optically thin tracers could be used in future studies to provide better estimates of the optical depths ($^{13}$CO(1-0) for example; see \citealt{weiss2005}).

\subsection{Shocked Gas}
\label{fe2}
\begin{figure*}
\includegraphics[width = \textwidth]{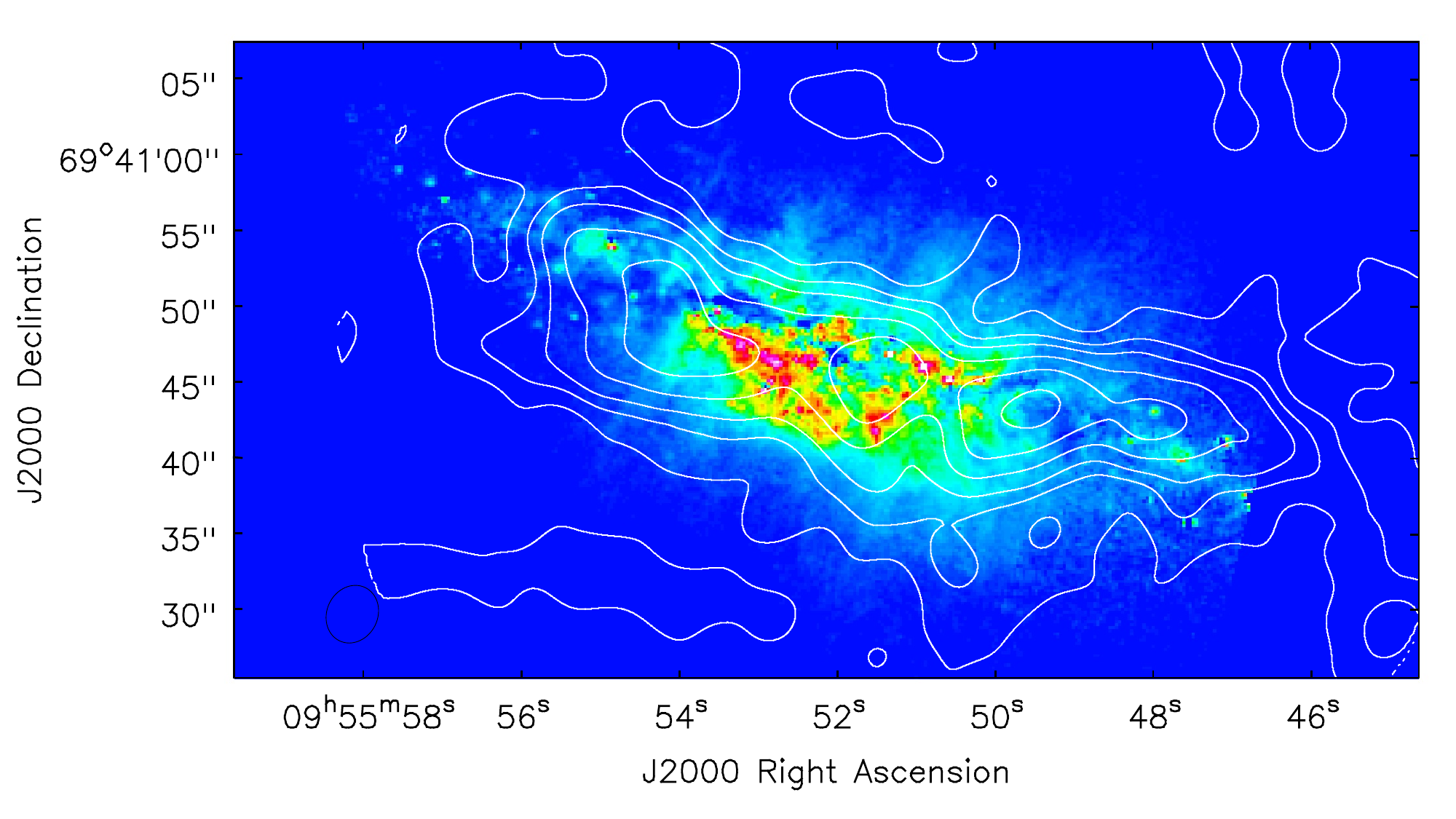}
\caption{HST [Fe~{\sc II}] emission map, overlaid with white \cotwo contours from \autoref{fig:mom0}. Knots of [Fe~{\sc II}] emission are near the intersection of S2 (upper left portion of the disc), and along the starburst driven bubble.}
\label{fig:fe2}
\end{figure*}

[Fe~{\sc II}] is a forbidden transition, and requires densities greater than 10$^4$~cm$^{-3}$ to populate the upper energy state \citep{mouri}. Fast (>100~\kmsp) shocks create high temperatures and densities that collisionally excite Fe~{\sc II}, producing [Fe~{\sc II}] emission in the infrared \citep{mouri}. [Fe~{\sc II}] is less susceptible to dust extinction than optical shock tracers (like [N~{\sc II}]) because of the lower dust attenuation in the infrared. Below, we primarily use these favorable characteristics to determine whether the molecular gas is shocked where S2 intersects the disc.

\autoref{fig:fe2} shows the [Fe~{\sc II}] emission map. The [Fe~{\sc II}] emission strongly peaks near the center of M~82, but the emission also traces the increased density enhancements and depressed temperatures of the super-bubble seen in the $-85$~\kms channel of \autoref{fig:chanden}. Intriguingly, there are knots of emission near the intersection of S2 and the disc (see the green X in \autoref{fig:mom0} and \autoref{fig:properties} for the approximate location of this feature).

\section{DISCUSSION}
\label{discussion}

Here we discuss three intriguing physical regions in the CO observations that probe the three phases of the baryon cycle: (1) the disc (\autoref{disc}) (2) the expanding super-bubble (\autoref{bubble}), and (3) the base of the S2 molecular streamer (\autoref{s2}). In each section we describe the region, its derived physical properties, how these properties compare to previous observations, and the mass of the feature. Finally, in \autoref{baryon} we discuss how the three areas illustrate the baryon cycle within M~82. 

\subsection{The Disc}
\label{disc}

The molecular disc of M~82 is well studied \citep{weiss99, mao2000, matsushita2000, matsushita2005, weiss2001, walter, weiss2005, keto05, salak13, leroy15}, and here we use our observations to calculate the molecular gas mass of the disc. We define the disc as the area within the 15$\sigma$ contours of the \cotwo intensity map. We make this distinction to avoid contributions from the super-bubble and the molecular streamers, but it affects comparisons with other studies. To compare the total mass directly with previous studies, we also calculate the total mass within the 450~K~km~s$^{-1}$ ($\approx$3$\sigma$) contours of the \coone map, which accounts for all of the observed CO.

The disc has a mean temperature of $104~\pm~36$~K, and log($n_\text{H2}$[cm$^{-3}$]) of $2.6\pm0.5$~dex. These values  are consistent with values from \citet{weiss2001} who use the IRAM Plateau de Bure Interferometer to derive mean temperatures and densities of $125\pm50$~K and  $3.4\pm0.5$~dex, respectively.

We calculate the total H$_2$ mass in two ways: (1) by converting the \coone intensity into a total molecular mass using an $X_\mathrm{CO}$ factor, and (2) by using the CO column densities ($N_\mathrm{CO}$) from the RADEX calculations. Method 1 is primarily used to compare with previous results which use a constant $X_\mathrm{CO}$ \citep{matsushita2000, weiss2001, walter}, while the second method uses all three transitions and the radiative transfer analysis to calculate the total H$_2$ mass. We focus on the results from method 2 while discussing the baryon cycle because it incorporates more of the known physics.

\begin{figure}
\includegraphics[width = 0.5\textwidth]{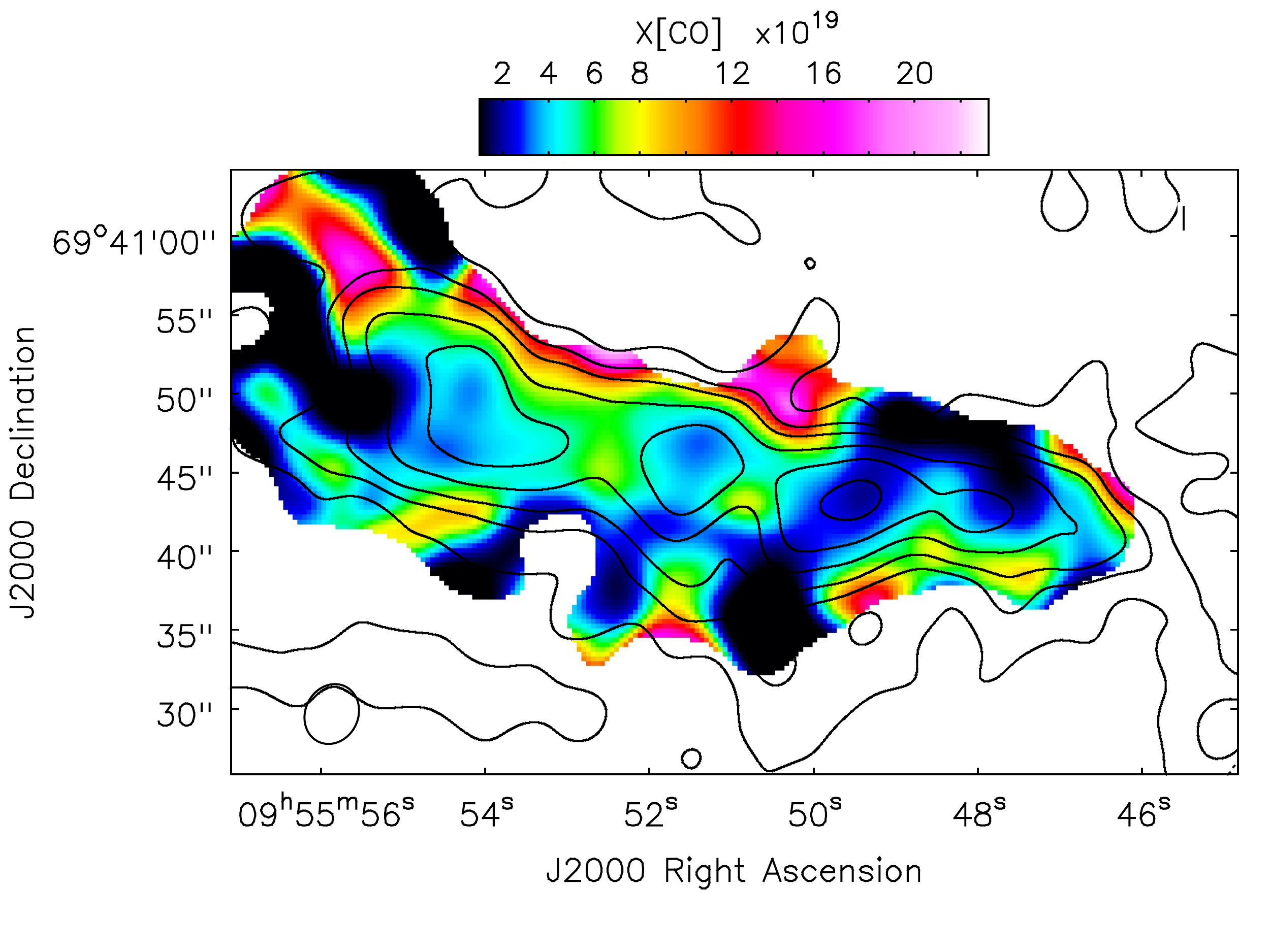}
\caption{Map of the $X_\mathrm{CO}$ factor (in units of 10$^{19}$~cm$^{-2}$/(K km s$^{-1}$)), as calculated using the RADEX models and \autoref{eq:xco} \citep{sakamoto1999}. The median $X_\mathrm{CO}$ factor in the disc is 4$\times$10$^{19}$~cm$^{-2}$(K km s$^{-1}$)$^{-1}$, nearly an order of magnitude lower than the Galactic $X_\mathrm{CO}$ value. The contours are the \cotwo zeroth moment map from \autoref{fig:mom0}, and the beam size(68~pc$ \times$ 58~pc) is given in the lower left. }
\label{fig:X}
\end{figure}

\begin{deluxetable*}{cccccccc}
\tablewidth{0pt}
\tablecaption{Mass Calculations}
\tablehead{
\colhead{(1)} &
\colhead{(2)} & 
\colhead{(3)} & 
\colhead{(4)} &  
\colhead{(5)} &
\colhead{(6)} &
\colhead{(7)} &
\colhead{(8)}  \\
\colhead{Component} &
\colhead{$A$} & 
\colhead{<$I_{10}$>} & 
\colhead{log$_{10}$($N_\mathrm{CO}$)} &  
\colhead{log$_{10}$($M_{H2}$) 1} &
\colhead{log$_{10}$($M_{H2}$) 2} &
\colhead{$\nabla$} &
\colhead{$\dot{M}$} \\
\colhead{} &
\colhead{(pc$^{2}$)} & 
\colhead{(K km s$^{-1}$)} & 
\colhead{(cm$^{-2}$)} &  
\colhead{(M$_\odot$)} &
\colhead{(M$_\odot$)} &
\colhead{(10$^{-7}$yr$^{-1}$)} &
\colhead{(M$_\odot$ yr$^{-1}$)}
}
\startdata
Total & 395467 & 831 & 20.10 & 8.96 & 8.32 & -- & -- \\
Disc  & 137702 & 1199 & 20.30 & 8.67 & 8.06  & -- & -- \\
Bubble & 77368  & 503 & 19.68 & 8.04 & 7.19 & 10.6 & 17.1 \\
S2 & 5650 & 708 & 19.81 & 7.41 & 6.18& 9.03 &  1.4
\enddata
\tablecomments{Table of the values used to calculate the total mass of the individual components: the total field of view, the disc, the super-bubble, and the base of the S2 streamer. Column 2 gives the area of each feature  (using 17.5~pc per arcsecond). Column 3 gives the average \coone intensity of the region, and column 4 gives the $^{12}$CO column density, calculated with RADEX (see \autoref{prop}). Column 5 gives the total molecular mass from method 1 (see \autoref{eq:massx}), and Column 6 gives the total molecular mass using method 2 (see \autoref{eq:massco}). Column 7 gives the velocity gradient derived from the PV diagram (only for the bubble and S2). Column 8 gives the mass outflow (inflow) rate, calculated using the mass from method 2 (column 6). For these calculations we assume a CO-to-H$_{2}$ abundance ratio of $5 \times 10^{-5}$ (see \autoref{tab:masscalc}).}
\label{tab:mass}
\end{deluxetable*}

\begin{deluxetable}{ccc}
\tablewidth{0pt}
\tablecaption{Constants Used in Mass Calculations}
\tablehead{
\colhead{Constant} &
\colhead{Value} & 
\colhead{units}\\
}
\startdata
$\bar{m}$&  $1.96 \times 10^{-57}$ & M$_\odot$/H$_2$ \\
$X_{CO}$ &  $1.5 \times 10^{20}$ & cm$^{-2}(K~$km~s$^{-1})^{-1}$ \\
Z[CO] & $5 \times 10^{-5}$ & CO/H$_2$ \\
$\Delta$~V & 89 & \kmsp
\enddata
\tablecomments{Constants used to calculate masses.}
\label{tab:masscalc}
\end{deluxetable}

Method 1 calculates the total H$_2$ mass as
\begin{equation}
M = X_{CO} <I_{10}> A \bar{m}
\label{eq:massx}
\end{equation}
where $X_\mathrm{CO}$ converts the median \coone intensity ($<I_{10}>$) into a H$_2$ column density, $A$ is the area of the region (in cm$^2$), and $\bar{m}$ is the average mass per H$_2$ molecule (see \autoref{tab:masscalc}). To compare to previous studies \citep{matsushita2000, weiss2001, walter, keto05}, we use an $X_\mathrm{CO}$ value that is half the Milky Way value  \citep[][\autoref{tab:masscalc}]{solomon87}. Using Method 1, the total H$_2$ mass within the 3$\sigma$ \coone contours is $9 \times 10^8$~M$_\odot$  (see \autoref{tab:mass}), similar to the $7 \times 10^8$~M$_\odot$ \citet{weiss2001} and \citet{walter} calculate using the $X_\mathrm{CO}$ method.

Method 2 uses the RADEX derived $N_\mathrm{CO}$ to calculate the total H$_2$ mass as
\begin{equation}
M = \frac{N_{CO} A \bar{m}}{\Delta V Z[CO]}
\label{eq:massco}
\end{equation}
where $\Delta V$ is the FWHM (89~\kmsp) and $Z[CO]$ is the CO-to-H$_2$ abundance ratio \citep[see \autoref{tab:masscalc};][]{sakamoto1999}. The total molecular gas mass within the field of view is 2.1$\times10^8$~M$_\odot$ (see the first row of column 6 in \autoref{tab:mass}), similar to the $2.7 \times 10^8$~M$_\odot$ \citep{weiss2001} estimated using an LVG solution.

In the disc, there are $5 \times 10^8$~M$_\odot$ and $1 \times 10^8 M_\odot$ of H$_2$, using methods 1 and 2, respectively (see \autoref{tab:mass}). The two methods differ largely because the excitation of CO depends on the temperature and density of the gas. For example, the $X_\mathrm{CO}$ of an optically thick cloud in virial equilibrium scales as $T^{-1}n_{H2}^{1/2}$ \citep{maloney, weiss2001}, and high H$_2$ temperatures within the disc decrease the conversion factor \citep{weiss2001, bolatto, leroy15}.  Higher temperatures excite higher energy transitions (\cothree for example), causing the \coone transition to trace a lower fraction of the total H$_2$. This relation only holds for clouds in virial equilibrium, and is meant to illustrate possible ways to decrease $X_\mathrm{CO}$. \citet{keto05} use high resolution ORVO \cotwo observations and this $X_\mathrm{CO}$ value to show that the CO clouds are approximately in virial equilibrium, but it is unlikely that all of the diffuse gas (scales of kpc) in M~82 is in virial equilibrium. To illustrate how the $X_\mathrm{CO}$ value changes, we use equation 1 from \citet{sakamoto1999} to calculate the $X_\mathrm{CO}$ value within the disc as
\begin{equation}
X_{CO} = \frac{N_{CO}}{\Delta V~Z[CO] <I_{10}>}
\label{eq:xco}
\end{equation}
In \autoref{fig:X} we find that the median disc value is $4.0 \times 10^{19}$~cm$^{-2}$/(K km/s), roughly consistent with the $X_\mathrm{CO}$ factor derived by \citet{weiss2001} using the LVG method. However, this $X_\mathrm{CO}$ factor is 4 times smaller than the value assumed in method 1, and almost an order of magnitude lower than the Milky Way value. The total mass is easily over-estimated if a constant $X_\mathrm{CO}$ factor is assumed, especially in regions with high temperatures and densities.

In summary, we calculate the total H$_2$ mass in two ways: with a constant conversion factor and with our radiative transfer calculations (see \autoref{tab:mass} for the values). Both methods are consistent within a factor of 2 with previous studies. We calculate the $X_\mathrm{CO}$ factor within the disc and find that it is also consistent with other studies that use radiative transfer techniques, however it is up to a factor of 4 lower than the constant value typically assumed. We now compare the total H$_2$ mass within the disc to the mass of the starburst driven super-bubble.

\subsection{The Starburst Driven Bubble}
\label{bubble}

The next feature that we discuss is the molecular super-bubble. At the center of the molecular bubble is an evolved 10$^6$~M$_\odot$ super-star cluster dominated by red super-giants. An estimated $4 \times 10^3$ supernovae have exploded throughout the lifetime of the cluster \citep{matsushita2000}. This energy and momentum has created an X-ray emitting plasma \citep{griffiths, matsushita2005} that adiabatically expands along the minor axis of the galaxy, and accelerates the cold molecular gas to velocities that reach $-139$~\kmsp (see \autoref{fig:pv21}). The $-85$~\kms channel maps (middle left panels in \Cref{fig:chanden,fig:chantemp}) show that the super-bubble has two components: (1) a moderate density and temperature shell that traces the [Fe~{\sc II}] emission in \autoref{fig:fe2} and (2) a high temperature, low density flow. The temperature and log($n_\text{H2}$[cm$^{-3}$]) in the shell are $82\pm17$~K and  $2.64\pm0.18$~dex, while the warm flow has a temperature of $159\pm30$~K and log($n_\text{H2}$[cm$^{-3}$]) of $1.93\pm0.15$~dex, significantly warmer and more rarefied than the disc. The dense shell is coincident with the strong [Fe~{\sc II}] emission (\autoref{fig:fe2}), indicating that shocks have compressed the H$_2$ to form the dense shell. 

Similar to the molecular disc, we calculate the total mass of the bubble in two ways using \autoref{eq:massx}, \autoref{eq:massco}, and the values from \autoref{tab:mass}. To calculate the mass of the super-bubble we use the intensity and $N_\mathrm{CO}$ found within the $-85$~\kms channel. This channel maximizes the emission from the bubble, while minimizing the contributions from the disc (see \autoref{maps}). The total mass in the super-bubble is $1.1 \times 10^{8} M_\odot$ for method 1. Using a similar $X_\mathrm{CO}$ factor, \citet{matsushita2000} find a total molecular mass  of $1.8 \times 10^8$~M$_\odot$. Using method 2 we find a total H$_2$ mass of $1.5 \times 10^7 M_\odot$, 7\% of the total H$_2$ mass within the galaxy.

We calculate the mass outflow rate using the velocity gradient from the PV diagram (see \autoref{fig:pv21}). The bubble initially expands nearly isotropically because it is not pressure-confined by the disc, as shown by the shell-like structure in the position-velocity diagram. This isotropic expansion velocity corresponds to the velocity of the molecular outflow \citep{matsushita2000}. The velocity gradient ($\nabla$) of the bubble extends from $-8$\farcs10 to 0\farcs76, with velocities from $-10$ to $-139$~\kmsp. Using the conversion factor that 1\farcs0 is 17.5~pc, we calculate $\nabla$ as
\begin{equation}
\nabla = \frac{\Delta v}{\Delta D}
\end{equation}
with a value of $1.1 \times 10^{-6}$~yr$^{-1}$. The total mass outflow rate ($\dot{M}$) is given by:
\begin{equation}
\dot{M} = M \nabla
\end{equation}
for a $\dot{M}$  of 17 M$_\odot$ yr$^{-1}$, using the mass from method 2. We do not calculate $\dot{M}$ for method 1 because the discrepancies associated with a constant X$_\mathrm{CO}$ factor (see \autoref{disc}).

Interestingly, the molecular mass outflow rate is similar to the large-scale atomic and ionized mass outflow rate calculated using the [O~{\sc I}] and [C{\sc~II}] emission lines \citep{contursi}. However, the molecular, neutral, and ionized outflows are in quite different physical locations: the CO emission dominates the inner kpc of the outflow \citep{walter, salak13, leroy15}, while the neutral and ionized emission extends further into the outflow \citep{shopbell, leroy15}.  This suggests that the molecular outflow is the base of the galactic outflow. Initially the stellar energy and momentum creates a hot plasma the size of the star-cluster. This hot plasma expands adiabatically and encounters the surrounding dense molecular gas within the disc, shocking and accelerating the molecular gas into the observed shell structure. The hot plasma rapidly heats and dissociates the molecular gas, mixing the cold and hot gas.  While the hot outflow initially contains a negligible amount of mass, the added molecular gas significantly increases the mass of the outflow \citep{chevalier, heckman90, cooper, strickland09}. This \lq{}\lq{}mass-loading\rq{}\rq{} transports cold gas out of the disc and into the halo through the galactic outflow, where it is later observed by warmer tracers like [O~{\sc I}], [C{\sc~II}], H$\alpha$, and metal absorption lines \citep{heckman2000, veilleux05, contursi}.  Mass-conservation requires that the outflow rate at the base of the outflow (the molecular gas) is equal to the outflow rate at later times and larger distances (the neutral and ionized gas), which is consistent with the observations of the atomic and ionized gas.

X-ray emission probes the mass-loading because it arises in the interaction region between the hot plasma and the molecular gas \citep{strickland2000, cooper}. The mass-loading factor ($\eta = \dot{M}_o/SFR$)  measures the efficiency of the outflow, relative to the star formation rate. With a star formation rate of 13~$M_\odot$~yr$^{-1}$ \citep{forster03}, M~82 has a molecular mass-loading factor of 1.3. Models of the X-ray Fe and S line fluxes of M~82 find that the mass-loading factor must be between 1.0 and 2.8 \citep{strickland09}, consistent with the molecular gas mass-loading found here.

After the molecular gas has been ejected from the disc, what is the final location of the gas? Does the molecular gas have enough kinetic energy to overcome gravity, or will the gas fall back as a galactic fountain \citep{shapiro}? The two scenarios illustrate very different implications for the baryon cycle: recycled gas can eventually be retained as stars, while escaping gas will decrease the fraction of baryons relative to the dark matter. Using a conservative estimate that the escape velocity is three times the circular velocity \citep{heckman2000} and  a circular velocity of 136~\kms \citep{yun}, the escape velocity of M~82 is 408~\kmsp, significantly higher than the the CO outflow velocity of 139~\kms  from \autoref{fig:pv21}. However, dissociated clouds of CO can be accelerated by ram pressure, and the clouds then may approach the escape velocity. \citet{chisholminprep} find a shallow scaling relation between SFR and the centroid velocity of the Si~{\sc IV} absorption lines (a warm ionized gas tracer). Using 13~M$_\odot$~yr$^{-1}$, these scaling relations predict the Si~{\sc IV} velocity would be 222~\kmsp, implying that the molecular gas must accelerate 87~\kms during the dissociation and ionization process. However, this velocity is still below the escape velocity, and the molecular gas is likely recycled. This result echoes the finding from \citet{leroy15} where the radial density profile of the gas and dust are too steep to produce a galactic outflow. The steep density profiles more naturally suggest that M~82's molecular bubble ends as a galactic fountain rather than a large scale outflow.

\subsection{The Bases of the Molecular Streamers}
\label{s2}

\citet{walter} identify four "streamers" in the wide field CO images. These streamers are large scale structures that extend 2~kpc from the edge of the molecular disc \citep{salak13, leroy15}. While we detect S1 and S3 in the \cotwo and \cothree images, the \coone field of view only provides adequate analysis of the base of the S2 streamer (the northeastern streamer). However, S2 is the base of a large H~{\sc I} trailing tail thought to have formed from the interaction of M~82 and M~81 \citep{yun, yun94}. The S2 streamer has the highest H~{\sc I} column density of the observed streamers ($5.6 \times 10^{20}$~cm$^{-2}$), and an estimated H~{\sc I} mass of $6.4 \times 10^7 M_\odot$ \citep{yun}.

S2 is set apart from the disc as a moderately high density and temperature region in the full temperature and density plots (\autoref{fig:properties}). In the +25~\kms channel map (lower left panel of \autoref{fig:chantemp}), S2 has a log($n_\text{H2}$[cm$^{-3}$]) of 2.3~dex and a temperature of  128~K. The kinematics of S2 also distinguish it from the disc, with S2 blueshifted $-37$~\kms from the disc (see \autoref{fig:mom1}). 

Further, in \autoref{fe2} we find that there is a strong knot of [Fe~{\sc II}] at the intersection of S2 and the disc, demonstrating that the gas is shocked at their intersection. The [Fe~{\sc II}] is not the only shock tracer available. \citet{westmoquette2009b} find elevated [N~{\sc II}]/H$\alpha$ and [S~{\sc II}]/H$\alpha$ ratios near the base of S2, but caution that the elevated ratios could be due to Hydrogen absorption. Finally, \citet{garcia} find that the two main sources of SiO in M~82 are the super-bubble, and the \lq\lq{}chimney\rq\rq{}, a large extended filament out of the disc. Figure~2 of \citet{garcia} shows a third source of SiO emission northeast of the main chimney, at a position and velocity consistent with the intersection of S2 and the disc. SiO traces shocked gas because it is released after a shock destroys dust grains \citep{pintado}. Therefore, the available shock diagnostics indicate that there is a shock at the intersection of S2 and the disc.

The distinct kinematics of S2 suggest that S2 is either an inflowing molecular filament, or a molecular outflow. These two scenarios are distinguished by the geometries of the system: if S2 is on the far side, the blueshifted emission implies that the gas is flowing onto the disc; whereas if S2 is on the near side, the blueshifted emission implies the gas is flowing out of the disc. Three observations lead us to hypothesize that S2 is an inflowing filament. First, if S2 is a molecular outflow there should be a burst of star formation that drives the CO out of the disc. There is not strong 100~GHz \citep{matsushita2005}, H$\alpha$ \citep{westmoquette2009b}, Pa-$\alpha$  \citep{alonso}, nor X-ray emission \citep{griffiths} in this region, implying that there is not a large amount of star formation in the vicinity. Secondly, there are strong shock indicators at the intersection of S2 and the disc. Gas flowing into a rotating disc will shock, but CO outflowing from the disc would not. Third, using the measured $N_\mathrm{CO}$ column density along S2, the CO-to-H$_2$ conversion factor, and a Calzetti extinction law \citep{calzetti2000}, we would expect S2 to extinct the galaxy by an A$_V$ of 28 magnitudes, which is not observed in optical images of S2 \citep{m82image}. Therefore, we assume that S2 is on the far side of the galaxy, and that S2 is a molecular inflow. In fact, S2 is likely the base of the large-scale gaseous streamer observed in H~{\sc I} that is created through the tidal interactions between M~82 and M~81 \citep{yun, yun94}.

To calculate the molecular gas mass inflow rate, we use the area defined by the kinematically distinct region of the \cotwo velocity map (see the velocity map in \autoref{fig:mom1}, and \autoref{fig:mom0} for the blue outline of the region). The total molecular gas mass of the base of S2, using method 1 (\autoref{eq:massx}), is $3 \times 10^7 M_\odot$, while the molecular gas mass is $2 \times 10^6 M_\odot$ using method 2 (\autoref{eq:massco}). The velocity gradient is measured from the PV diagram, and found to be $9 \times 10^{-7}$~yr$^{-1}$, for a total molecular gas mass inflow rate of $1.4~ M_\odot$~yr$^{-1}$ (see \autoref{tab:mass}). 

Here, we only observe the {\it molecular} accretion, but there are other contributions to the total baryon cycle in M~82. M~82 gains and loses neutral and ionized gas, while dissociation, ionization, and recombination convert the molecular, neutral, and ionized gas into other phases. This leads to a fluid, and complicated, inflow rate. Furthermore, the accretion we probe is only filamentary, and mostly due to the tidal interaction with M~82's companions. Spherical accretion assuredly occurs in many of the phases. Unfortunately, spherical accretion is difficult to kinematically distinguish from disc gas because the velocities of both are set by the gravity of the galaxy. Therefore, the accretion rate we measure is a lower limit of both the molecular accretion rate and the total accretion rate.

\subsection{The Molecular Baryon Cycle}
\label{baryon}
In the past three sections we have outlined how gas is accreted onto (\autoref{s2}), resides within (\autoref{disc}), and ejected out (\autoref{bubble}) of the nearby starburst M~82. Additionally, baryons can be locked into stars through star formation, and M~82 is forming stars at a rate of 13~$M_\odot$~yr$^{-1}$ \citep{forster03}.  This traces the entire molecular baryon cycle.  To remain in a steady-state, M~82 requires a molecular inflow rate of  30~$M_\odot$~yr$^{-1}$ (SFR plus $\dot{M}$), but we only observe a molecular inflow rate of 1.4~$M_\odot$~yr$^{-1}$ from one  -- of the four known -- molecular streamers.  

\citet{yun} find that the S2 streamer contains 40\% of the observed H~{\sc I} mass in the streamers. If we assume that the H~{\sc I} is a proxy for the H$_2$ and that the other streamers accrete onto the disc at the same relative rate as S2, the total inferred molecular inflow rate is 3.5~M$_\odot$~yr$^{-1}$ (1.4~M$_\odot$~yr$^{-1}$/0.4). Using this accretion rate, we find that M~82 is consuming, and expelling molecular gas 9 times faster than it is acquiring molecular gas. The molecular baryon cycle in M~82 is running a deficit: if the current rates continue, M~82 will consume or expel all of the observed molecular gas in 7.8~Myr. With over $3 \times 10^8$~M$_\odot$ of H~{\sc I} in and around the galaxy \citep{yun}, converting H~{\sc I} into H$_2$ may extend this depletion timescale \citep{krumholz08, krumholz09}, but how efficiently H~{\sc I} is converted into H$_2$ is uncertain. Additionally, ionized gas can recombine and eventually cool to molecular gas, and spherical accretion of ionized gas may provide another source of molecular gas. However, if the molecular gas acquisition rate does not increase in the next 8~Myr, the star formation rate or the outflow rate must decrease, or M~82 will exhaust all of its molecular gas.

\section{CONCLUSION}

We present \cotwo and \cothree observations of M~82, a nearby starburst galaxy. We combine these measurements with previous \coonep, and [Fe~{\sc II}] emission lines to illustrate the physical properties of the molecular gas. We calculate the molecular gas mass of three previously identified features: the disc, an expanding super-bubble, and the base of a molecular streamer (named S2). Using the kinematics, shock tracers (SiO emission and [Fe~{\sc II}] emission), and optical extinction, we argue that S2 is an inflowing filament of molecular gas that is shocked as it encounters the disc. We then compare the star formation rate (13~$M_\odot~$yr$^{-1}$), molecular mass outflow rate (17~$M_\odot$~yr$^{-1}$), and molecular mass inflow rate (3.5~$M_\odot$~yr$ ^{-1}$) to find that the molecular gas is consumed more rapidly than it is replenished by the inflow. We conclude that the baryon cycle in M~82 is running a deficit: unless more molecular gas is acquired, the star formation rate or outflow rate must decrease in the next eight million years.

\section*{Acknowledgments}
We thank the referee, Professor Adam Leroy, for a very thoughtful and considerate review of the paper, that considerably improved the analysis and interpretation of the data.

We thank Sarah Wood from NRAO for her persistence and invaluable help with the reduction and manipulation of single dish data.

We thank Jay S. Gallagher for helpful discussions while composing the manuscript and analyzing the Hubble Space Telescope data.

The Submillimeter Array is a joint project between the Smithsonian Astrophysical Observatory and the Academia Sinica Institute of Astronomy and Astrophysics and is funded by the Smithsonian Institution and the Academia Sinica.
Some of the data presented in this paper were obtained from the Mikulski Archive for Space Telescopes (MAST). STScI is operated by the Association of Universities for Research in Astronomy, Inc., under NASA contract NAS5-26555. Support for MAST for non-HST data is provided by the NASA Office of Space Science via grant NNX09AF08G and by other grants and contracts.

This material is based upon work supported by the National Science Foundation through the East Asia South Pacific Institute (EAPSI) program under Grant No. (1310907).

SM is supported by the National Science Council (NSC) and the Ministry of Science and Technology (MoST) of Taiwan, NSC 100-2112-M-001-006-MY3 and MoST 103-2112-M-001-032-MY3.

\bibliographystyle{apj}
\bibliography{m82bubble}

\end{document}